\documentclass[preprint]{aastex}

\shorttitle{Measuring the speed of light with ultra-compact radio
quasars} \shortauthors{Cao et al.}

\begin{document}


\title{Measuring the speed of light with ultra-compact radio quasars}


\author{Shuo Cao}
\affil{Department of Astronomy, Beijing Normal University,
    Beijing 100875, China}

\author{Marek Biesiada}
\affil{Department of Astronomy, Beijing Normal University,
    Beijing 100875, China; \\
    Department of Astrophysics and Cosmology, Institute of Physics, University of Silesia, Uniwersytecka 4, 40-007, Katowice, Poland}

\author{John Jackson}
\affil{Department of Astronomy, Beijing Normal University, Beijing
100875, China}

\author{Xiaogang Zheng}
\affil{Department of Astronomy, Beijing Normal University, Beijing
100875, China}

\author{Yuhang Zhao}
\affil{Department of Astronomy, Beijing Normal University, Beijing
100875, China}

\and

\author{Zong-Hong Zhu}
\affil{Department of Astronomy, Beijing Normal University,
    Beijing 100875, China}
\email{zhuzh@bnu.edu.cn}



\begin{abstract}

In this paper, based on a 2.29 GHz VLBI all-sky survey of 613
milliarcsecond ultra-compact radio sources with $0.0035<z<3.787$, we
describe a method of identifying the sub-sample which can serve as
individual standard rulers in cosmology.  If the linear size of the
compact structure is assumed to depend on source luminosity and
redshift as $l_m=l L^\beta (1+z)^n$, only intermediate-luminosity
quasars ($10^{27}$ W/Hz$<L<$ $10^{28}$ W/Hz) show negligible
dependence ($|n|\simeq 10^{-3}$, $|\beta|\simeq 10^{-4}$), and thus
represent a population of such rulers with fixed characteristic
length $l=11.42$ pc. With a sample of 120 such sources covering the
redshift range $0.46<z<2.80$, we confirm the existence of dark
energy in the Universe with high significance under the
assumption of a flat universe, and obtain stringent constraints on
both the matter density $\Omega_m=0.323^{+0.245}_{-0.145}$ and the
Hubble constant $H_0=66.30^{+7.00}_{-8.50}$ km sec$^{-1}$
Mpc$^{-1}$. Finally, with the angular diameter distances $D_A$
measured for quasars extending to high redshifts ($z\sim 3.0$), we
reconstruct the $D_A(z)$ function using the technique of Gaussian
processes. This allows us to identify the redshift corresponding to
the maximum of the $D_A(z)$ function: $z_m=1.70$ and the
corresponding angular diameter distance $D_A(z_m)=1719.01\pm43.46$
Mpc. Similar reconstruction of the expansion rate function $H(z)$
based on the data from cosmic chronometers and BAO gives us
$H(z_m)=176.77\pm6.11$ km sec$^{-1}$ Mpc$^{-1}$. These measurements
are used to estimate the speed of light: $c=3.039(\pm0.180)\times
10^5$ km/s. This is the first measurement of the speed of light in a
cosmological setting referring to the distant past.

\end{abstract}





\section{Introduction}\label{sec:introduction}

Advances in cosmology over recent decades have been accompanied by
intensive searches for reliable standard rulers in the Universe.
Recently, attention of has been focused on large comoving length
scales revealed in the baryon acoustic oscillations (BAO). The BAO
peak  location is universally recognized as a fixed comoving ruler
of about 105$h^{-1}$ Mpc (where $h$ is the Hubble constant $H_0$
expressed in units of $100 \; km \, s^{-1} Mpc^{-1}$). However, the
so-called fitting problem \citep{Ellis87} still remains a challenge
for BAO peak location as a standard ruler. In particular, the
environmental dependence of the BAO location has recently been
detected by \citet{Roukema2015,Roukema2016}. Moreover,
\citet{Ding2015} and \citet{Zheng2016} pointed out a noticeable
systematic difference between $H(z)$ measurements based on BAO and
those obtained with differential aging techniques. Recently efforts
have also been made to explore the sizes of galaxy clusters  at
different redshifts, by using radio observations of the
Sunyaev-Zeldovich effect together with X-ray emission
\citep{Filippis05,Boname06}. In similar spirit, powerful radio
sources have served as a special population to test the redshift -
angular size relation for extended FRIIb galaxies \citep{Daly03},
radio galaxies \citep{Guerra00}, and radio loud quasars
\citep{Buchalter98}. More promising candidates in this context are
ultra-compact structures in radio sources (especially in quasars
that can be observed up to very high redshifts), with milliarcsecond
angular sizes measured by the very-long-baseline interferometry
(VLBI) \citep{Kellermann93,Gurvits94}; for details regarding the
respective definitions of angular size see \textit{Data}. Possible
cosmological applications of such compact radio sources as standard
rulers  have been extensively discussed in the literature
\citep{Jackson97,Vishwakarma01,Lima02,Zhu02,Chen03}.

The idealized picture of standard rulers is that of a population of objects whose linear sizes $l_m$ are the same and either constant or change with redshift in a well-known way. Of course, real objects like radio sources do not meet such stringent criteria. More realistically we hope for a population whose \textit{mean} linear size satisfies these criteria.
The extended radio
sources whose linear extent is $\sim 100$ kpc are expected to depend
strongly on the redshift evolution of the intergalactic medium.
Compact radio sources depend more on the properties of the central
engine (accreting central black hole) controlled by a limited number
of parameters (like mass, accretion rate, magnetic field), and the
ages of their jets are short compared with the age of the Universe.
Consequently it is reasonable to suppose that such sources are much
less affected by evolutionary processes. However, they are
inherently a mixed population of AGNs (quasars, BL Lac objects
etc.). There were suggestions \citep{Gurvits99,Vishwakarma01} that
the exclusion of sources with extreme spectral indices and low
luminosities might alleviate the dependence of $l_m$ on the source
luminosity and redshift. More recently, Cao et al. (2015) reexamined
the same data in the framework of the $\Lambda$CDM cosmological
model assuming parametrized dependence of $l_m$ on luminosity and
redshift (see Eq.(2) below). The general result was that whereas the
evolution of $l_m$ with redshift is small, its dependence on
luminosity can be substantial. We have divided the full sample into
various sub-samples according to optical counterpart and luminosity,
with view to finding which sub-sample performs best in the role of
standard rulers; quantitative details differ between the
sub-samples. In this paper we restrict our study to quasars and
consider three luminosity ranges separately. Full details of this
procedure will be the subject of a separate paper.

Here we focus on a very specific cosmological application of
standard rulers. Namely, we reconstruct the angular diameter
distance as a function of redshift up to $z \simeq 2.5$ and identify
the redshift $z_m$ corresponding to the maximum of the $D_A(z)$
function. According to \citet{Salzano15}, this allows us to
calculate the speed of light $c$, which is the
first empirical assessment of the speed of light at an epoch much
earlier than the present time.

\section{Observational data}\label{sec:data}

Cosmological parameter estimation from the angular-size/redshift
relation for ultra-compact radio sources was first discussed by
\citet{Kellermann93}, using a sample of 82 sources with 5 GHz VLBI
contour maps; these images show a compact core with outlying weaker
components, angular size being defined as the distance between the
core and the most distant component with peak brightness greater
than or equal to 2\% of that of the core. \citet{Gurvits99} extended
this work using a much larger sample, 330 VLBI images, taken largely
from the Caltech-Jodrell Bank 5 GHz survey \citep{Taylor96}. Their
main conclusion was that their 5 GHz VLBI data are consistent with
standard FLRW cosmologies with $0<\Omega_\Lambda<0.5$ and
$\Omega_\Lambda=0$. Their work has formed the basis of many
subsequent investigations, early examples being
\citet{Lima02,Zhu02,Chen03}.

The data used in the present paper were derived from an ancient 2.29
GHz VLBI survey undertaken by \citet{Preston85} (hereafter called
P85).  By employing a world-wide array of dishes forming an
interferometric system with an effective baseline of about $8\times
10^7$ wavelengths, this survey succeeded in detecting interference
fringes from 917 radio sources out of a list of 1398 candidates
selected mainly from the Parkes survey \citep{Bolton79}. The results
of this survey were utilized initially to provide a very accurate
VLBI celestial reference frame, improving precision by at least an
order of magnitude, compared with earlier stellar frames. An
additional expectation was that the catalog would be ``used in
statistical studies of radio-source properties and cosmological
models" \citep{Preston85}. It should be noted that P85 does not list
angular sizes explicitly; however, total flux density and correlated
flux density (fringe amplitude) are listed; the ratio of these two
quantities is the visibility modulus $\Gamma$, which defines a
characteristic angular size
\begin{equation}
\theta={2\sqrt{-\ln\Gamma \ln 2} \over \pi B} \label{thetaG}
\end{equation}
where $B$ is the interferometer baseline, measured in wavelengths
\citep{Thompson86,Gurvits94}. It is argued in \citet{Jackson04} that
this size represents that of the core, rather than the angular
distance between the latter and a distant weak component. In this
way \citet{Gurvits94} was able to construct an angular-size/redshift
diagram for 337 sources with known redshifts; those with $z<0.5$
were ignored, and using just the high-redshift data (258 objects)
found marginal support for a low-density CDM cosmological model, but
considered only those with $\Omega_\Lambda=0$. Using exactly the
same data set, \citet{Jackson97} extended this work to the full
$\Omega_m-\Omega_\Lambda$ plane, and concluded that the then
canonical flat CDM model is ruled out at the 98.5\% confidence
level, and that the best values are $\Omega_m=0.2,
\Omega_\Lambda=0.8$ if the Universe is spatially flat, later refined
to $\Omega_m=0.24^{+0.09}_{-0.07}$ \citep{Jackson04}. This work was
extended further by \citet{Jackson06}, who updated the
P85/\citet{Gurvits94} sample with respect to redshift, to include a
total of 613 objects with redshifts $0.0035\leq z\leq 3.787$. This
extended data set is the one used in the present paper. Following
\citet{Gurvits99}, we exclude sources with inverted spectra and
those with steep spectra: we apply an inclusion criterion on
spectral index ($-0.38\leq \alpha\leq 0.18$), leaving 240 sources
with relatively flat spectra. Appropriate information about all the
240 sources can be extracted from the source data set
\citep{Jackson06}. For each object, the corresponding optical
counterpart and spectral index between 2700 and 5000 MHz can be
found in P85. In order to enhance accuracy, in all cases we have
derived the values for these quantities from the original
references. We note that this final sample, which covers the
redshift range $0.014<z<3.707$, contains 181 sources identified as
quasars ($0.02<z<3.707$). By restricting the range of spectral
indices and concentrating on milliarcsecond radio structures in
quasars, we are better able to restrict the dispersion in intrinsic
size in our analysis \citep{Kellermann93,Gurvits99}.

If we observe a given millarcsecond source at different reception
frequencies $\nu_r$, it is well-known that its observed size falls
as the frequency increases, being proportional to $\nu_r^{-1}$
\citep{Marscher80,O'Sullivan09}. At first sight an expectation would
be that angular sizes in corresponding angular-size/redshift
diagrams should fall by an additional factor $(1+z)^{-1}$, because
the rest-frame emitted frequency $\nu_e$ has to be greater than the
received frequency $\nu_r$ by a factor of $(1+z)$:
$\nu_e=(1+z)\nu_r$, thus masking other cosmological effects.
However, observations show behaviour which is compatible with
conventional cosmologies. The reason is almost certainly that there
is a selection effect which is operating in our favour in this
context; we are dealing with an ensemble of objects which may be
intrinsically similar in their respective rest-frames, but appear to
be very different in our frame. The underlying population consists
of compact symmetric objects \citep{Wilkinson94}, each comprising a
central low-luminosity nucleus straddled by two oppositely-directed
jets. Ultracompact objects are identified as cases in which the jets
are moving relativistically and are close to the line of sight, when
Doppler boosting allows just that component which is moving towards
the observer to be observed, giving rise to the core-jet structure
observed in typical VLBI images, see for example
\citet{Pushkarev12}; the core is believed to be the base of the jet,
rather than the nucleus \citep{Blandford79}. Those jets which are
closest to the line of sight appear to be the brightest.  The radio
sample is flux-limited, so that as $z$ increases a larger Doppler
boosting factor D is require \citep{Dabrowski95}; it turns out that
the ratio $D/(1 + z)$  is approximately fixed, so that the
rest-frame emitted frequency $(1 + z)\nu_r/D$ is also fixed.  See
\citet{Jackson04} for mathematical and astrophysical details.  Note
that this ratio is not necessarily unity, but the fact that it is
fixed means that the interferometric angular sizes upon which this
work is based are fit for purpose. This is another reason for
ignoring sources with $z<0.5$, which appear to be non-relativistic.
The above picture is based upon the unified model of active galactic
nuclei and quasars \citep{Antonucci93, Antonucci15}.

\section{Method}\label{sec:method}

With respect to the milliarcsecond radio-sources, our procedure
follows that given in \citet{Sandage88}. For a standard rod, the
angular size at redshift $z$ can be written as
\begin{equation}
\theta(z)= \frac{l_m}{D_A(z)} \label{theta}
\end{equation}
where $l_m$ is the intrinsic metric linear size and $D_A$ represents
the angular diameter distance. Moreover, we include two parameters
$\beta$ and $n$ to respectively consider the ``angular size -
redshift" and ``angular size - luminosity" relations. Our procedure
follows the approach first proposed in \citet{Gurvits94} and later
investigated in \citet{Gurvits99}:
\begin{equation} \label{lm}
l_m=lL^\beta(1+z)^n\,
\end{equation}
where $l$ is the linear size scaling factor representing the
apparent distribution of radio brightness within the core, $L$ is
the luminosity normalized by $L_0=10^{28}$WHz$^{-1}$. In our
analysis, it has been estimated from the measured flux density
$S_{obs}$ and the spectral index $\alpha$ according to the formula:
$L=4 \pi {D_{A}}^{2}S_{\mathrm{obs}}/(1+z)^{\alpha-3}$. In the
latter we have used the so-called distance duality relation
connecting luminosity distance $D_L(z)$ with $D_A(z)$.
This fundamental relation has triggered many studies in observational
cosmology \citep{Cao11a,Cao11b,Cao16}.
The $\beta$ parameter, which
captures the dependence of the linear size on source luminosity, is
highly dependent on the physics of compact radio emitting regions
\citep{Gurvits99,Jackson04}. Besides cosmological evolution of the
linear size with redshift, the parameter $n$ may also characterize
the dependence of the linear size on the emitted frequency, as well
as image blurring due to scattering in the propagation medium.
Meanwhile, according to a recent analysis carried out by
\citet{Koay14}, ground-based interferometers could hardly detect the
scatter broadening of radio compact sources due to ionized
intergalactic medium (IGM), which can only be detected by Space VLBI
with baselines of 350,000 km at frequencies below 800 MHz.
Considering that all VLBI images for our sample were observed at a
frequency of 2.29 GHz, the effect of scattering in the propagation
medium is not important in the present analysis \citep{Gurvits99}.

It is obvious that, in combination with a reliable theoretical
expression for $D_A$, constraints on $\beta$, and $n$ will help us
to differentiate between sub-samples and determine standard rods
fulfilling the criteria of $\beta = n = 0$. To achieve this goal, we
assume flat $\Lambda$CDM model and use the cosmological parameters
given by recent \textit{Planck} Collaboration:
$\Omega_m=0.315\pm0.017$ and $h=0.673\pm0.012$ \citep{Planck1}.
Using routines available in the public CosmoMC package, we perform
Monte Carlo simulations of the posterior likelihood ${\cal L} \sim
\exp{(- \chi^2 / 2)}$ for $l$, $\beta$, $n$, in order to obtain
their best-fit values, and determine the corresponding confidence
regions for all three parameters characterizing compact radio
structures. In computing $\chi^2$ we have assumed 10\% uncertainties
in the observed angular sizes, which in effect allows for both
measurement errors and the intrinsic spread in linear sizes.

We should stress that, after identifying the best standard ruler
population, the cosmological-model-independent method will then be
applied to determine the linear size of this standard rod. Next,
using this value, the angular diameter distance function $D_A(z)$
will be further investigated, searching for the redshift $z_m$
corresponding its maximum and the value of the speed of light at
this redshift. It has been known for a long time \citep{Weinberg72}
that theoretically, angular diameter distance $D_A$ will always
ascend to a maximum value at a certain redshift $z_m$ and tend to
decline at higher redshifts. However, it is very difficult to check
this redshift evolution from astrophysical observations due to the
so-called ``redshift desert''. Moreover, the exact value for $z_{m}$
is highly dependent on the cosmological model adopted. In the
framework of a flat Friedmann universe with no cosmological
constant, the maximum redshift is fixed at $z_{m} = 1.25$. Numerical
simulations by \citet{Salzano15}, who generated $\sim 10^{4}$ random
cosmological models using the recent \textit{Planck+WMAP+highL+BAO}
best fitted cosmological parameters and their uncertainties,
demonstrated that 95\% of cases cover the range $z_{m}=[1.4,1.8]$.
In the present paper, considering the redshift range of our sample,
especially that of the high-redshift quasars we are able to trace
the evolution of $D_A(z)$ better and hence obtain more stringent
constraint on $z_m$. This is particularly interesting and important
because the necessary condition for the extremum $\partial D_{A}(z)/
\partial z = 0$ in the flat Universe leads to the formula for the
speed of light:
\begin{equation}
c=D_{A}(z_{m})H(z_{m}),
\end{equation}
where $H(z)$ is the Hubble parameter at redshift $z$
\citep{Salzano15}. Therefore, having the real data on $D_{A}(z)$ and
$H(z)$ one would be able to estimate the value of $c$ using extragalactic objects. This is the first such determination, since the
previous papers on this subject \citep{Salzano15,Salzano16},
although discussing the idea, referred only to simulated BAO data
representative of future surveys like SKA or Euclid.

\begin{table*}
\caption{\label{tab:result} Summary of constraints on the
linear size parameters obtained with different quasar sub-samples
(see text for the details). }
\begin{center}{\scriptsize
\begin{tabular}{|l|c|c|c|}\hline\hline

Quasar sub-samples (Data number)     & $l$ [pc]        & $\beta$   &  $n$ \\
\hline

Low-luminosity quasars (N=30)   & $48.07\pm18.42$     & $0.503\pm0.060$  & $-0.694\pm0.661$  \\
$[L<$ $10^{27}$ W/Hz$]$ & & & \\
\hline

High-luminosity quasars (N=31)   & $18.39\pm3.52$    & $0.300\pm0.104$  & $-0.579\pm0.175$  \\
$[L>$ $10^{28}$ W/Hz$]$ & & & \\
\hline

Intermediate-luminosity quasars (N=120)  & $11.19\pm1.64$    & $0.0001\pm0.052$  & $0.007\pm0.142$  \\
$[10^{27}$ W/Hz$<L<$ $10^{28}$ W/Hz$]$ &&&  \\
 \hline

Low-luminosity and high-redshift quasars (N=13)      & $22.01\pm16.75$    & $0.020\pm0.225$  & $-1.23^{+1.01}_{-0.87}$  \\
$[10^{26}$ W/Hz$<L<$ $10^{27}$ W/Hz; $z>0.50]$ & & & \\
\hline

High-luminosity and high-redshift quasars (N=31) & $18.39\pm3.52$    & $0.300\pm0.104$  & $-0.579\pm0.175$  \\
$[10^{28}$ W/Hz$<L<$ $10^{29}$ W/Hz; $z>0.50]$ &&&  \\
\hline

Combined High and Intermediate-luminosity quasars (N=151) & $14.76 \pm 1.53$    & $0.161 \pm 0.025$  & $0.229 \pm 0.104$  \\
$[L>$ $10^{27}$ W/Hz$]$  &&&  \\
\hline \hline

Intermediate-luminosity quasars (N=120) & $11.04\pm1.62$    & $-0.0024\pm0.050$  & $0.015\pm0.143$  \\
$[$Cosmology-independent method I$]$ &&& \\
\hline

Intermediate-luminosity quasars (N=120) & $10.86\pm1.58$    & $-0.0034\pm0.050$  & $0.027\pm0.143$  \\
$[$Cosmology-independent method II$]$ &&& \\
\hline

\end{tabular} }
\end{center}
\end{table*}

\section{Results and discussion}\label{sec:results}

\begin{figure*}
\begin{center}
\includegraphics[scale=0.45]{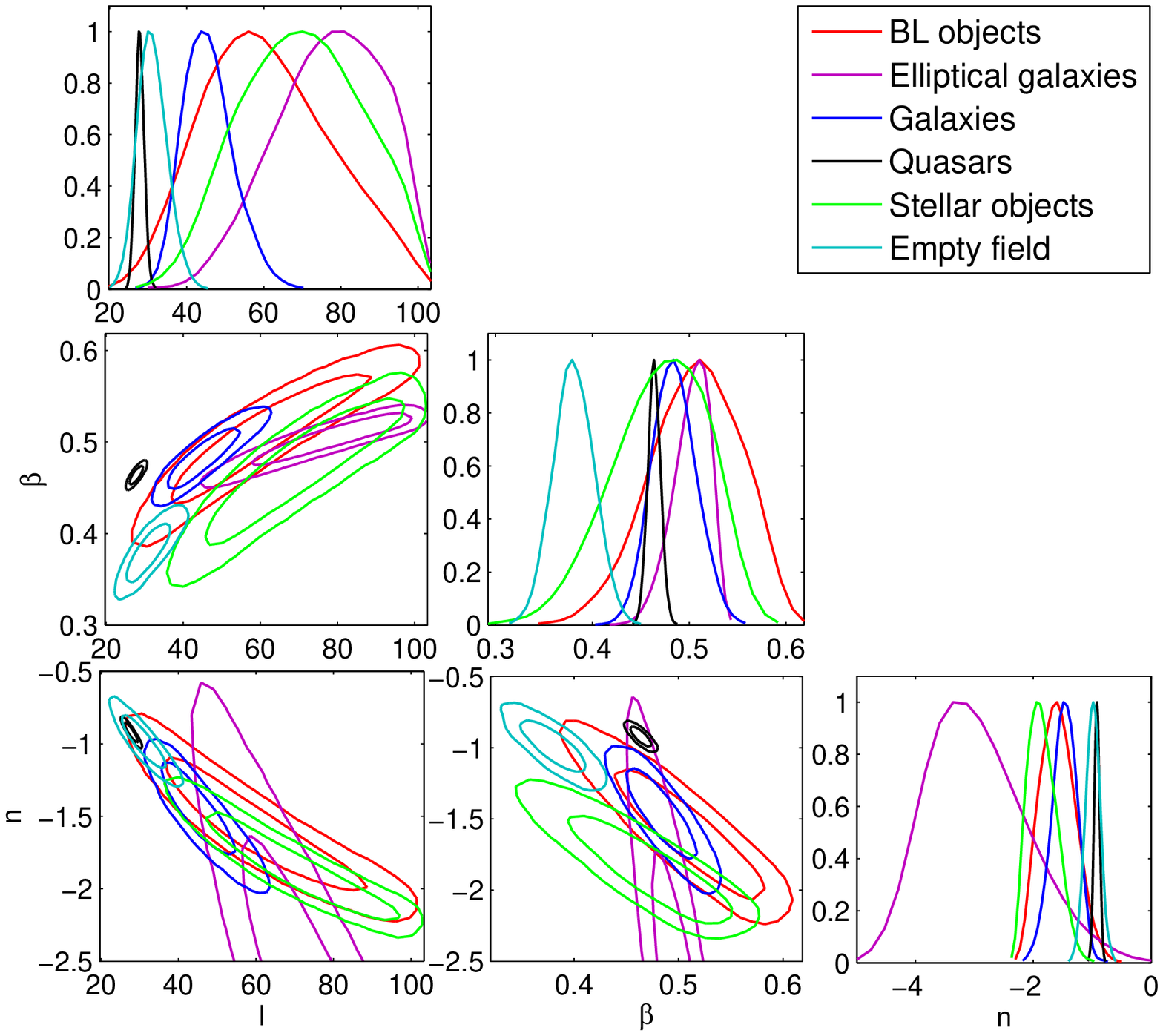} \includegraphics[scale=0.45]{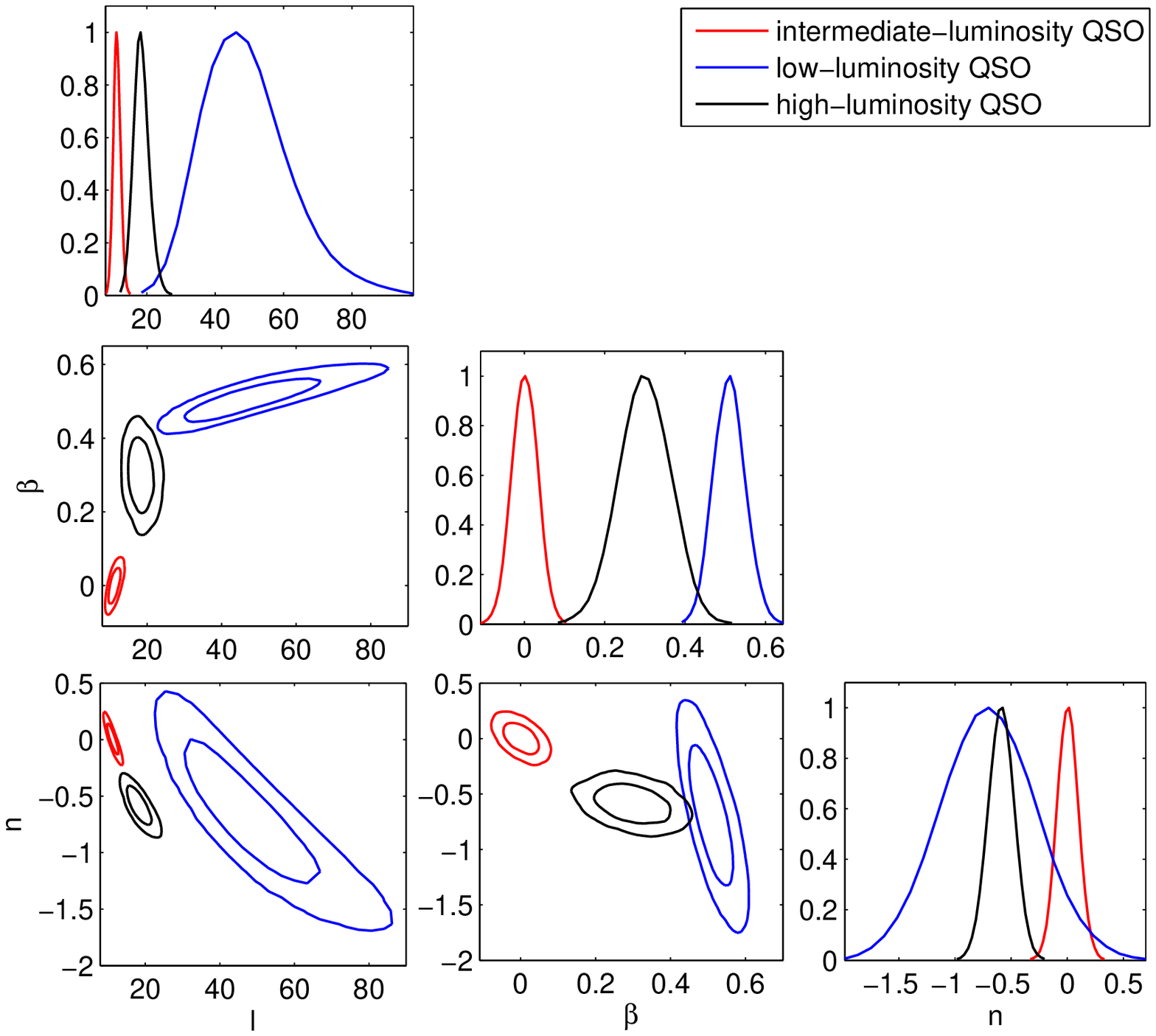}
\end{center}
\caption{Left panel shows the constraints on compact source
parameters obtained from samples with different optical
counterparts. Right panel refers to the quasars with different
luminosities: low luminosity $L < 10^{27} W/Hz$; intermediate
luminosity $10^{27} W/Hz<L< 10^{28} W/Hz$ and high luminosity $L>
10^{28} W/Hz$. Only the intermediate luminosity quasars display
negligible dependence of their linear size parameter on the source
luminosity and redshift ($|n|\simeq 10^{-3}$ and $|\beta|\simeq
10^{-4}$). }
 \label{fig1}
\end{figure*}

Let us start with the constraints on the parameters ($l$,
$\beta$, $n$) obtained for different samples corresponding to different
optical counterparts \footnote{In order to guarantee the appropriate
size of each sub-sample, we decided to use, for the purpose of
fitting, the full $n=613$ sample without the spectral index
criterion.}. Fig.~\ref{fig1} shows the corresponding $68\%$ and
$95\%$ confidence regions and marginalized distributions for each
parameter. Obviously, (as one can see on the left panel of the Fig.~\ref{fig1} ) fitting results for the $n$ and $\beta$
parameters for quasars are more stringent and quite different from
those derived with other compact radio sources, which might reveal
the different physical processes governing the radio emission of
compact structures in quasars \citep{Cao15}.

\begin{figure*}
\begin{center}
\includegraphics[angle=270,width=15cm]{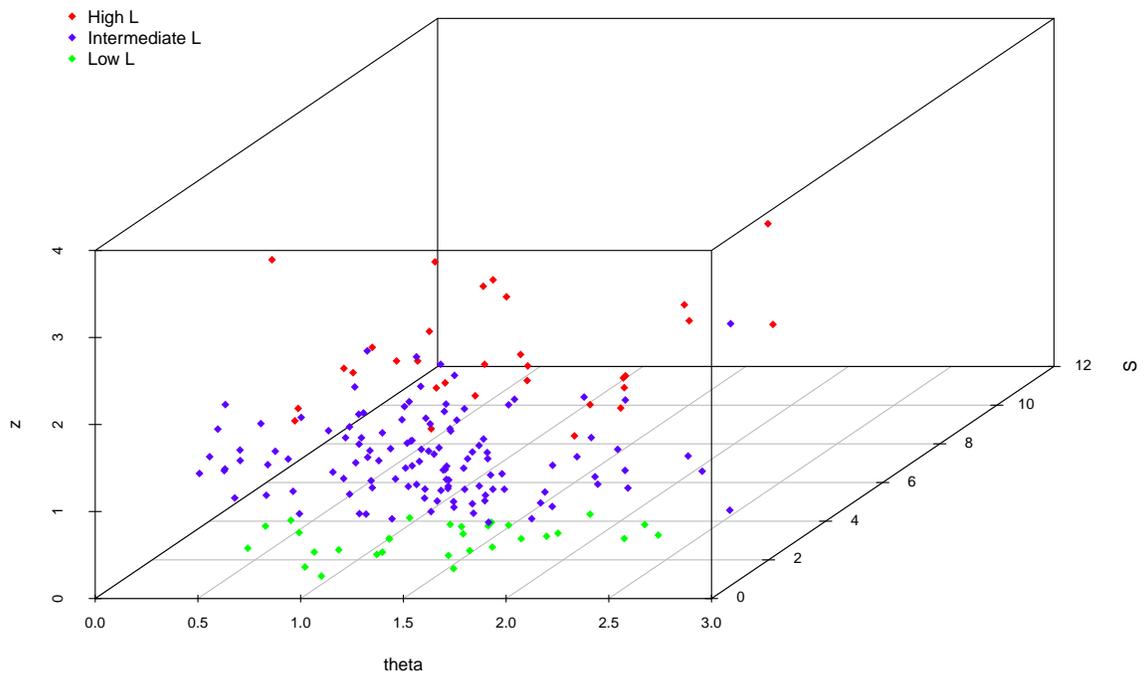}
\end{center}
\caption{ Scatter plot of three quasar sub-samples in the space of observable quantities: the angular size $\theta$ in $[mas]$, the observed flux $S$ at $2.29\;GHz$ in $[Jy]$, and the redshift $z$. Sub-samples selected by the luminosity criterion are marked with different colors. }
 \label{fig2}
\end{figure*}

Next, we consider only radio quasars, which choice is supported by
the following arguments. Firstly, quasars constitute the most
important part of our full radio sample. Secondly, the potential of
a quasar to act as a standard cosmological rod was already noticed
in \citet{Cao15}. However, the range of the parameter $\beta$
suggests that the linear sizes of quasars are significantly affected
by their luminosity. Therefore, we further divide the full $N=181$
quasar sample into three sub-samples, according to their luminosity
$L$. Quasars with luminosity smaller than $10^{27}$ W/Hz are
classified as relatively low-luminosity quasars, those with
luminosity larger than $10^{28}$ W/Hz are classified as relatively
high-luminosity quasars, and intermediate-luminosity quasars are
defined as having $10^{27}$ W/Hz$<L<$ $10^{28}$ W/Hz. The
intermediate-luminosity range corresponds to the interquartile range
of luminosity distribution in our total sample of quasars. Since the
determination of luminosity involves cosmological model through
$D_A(z)$, we have cross-checked our sample selection by using two
different approaches: using a fiducial cosmology (flat $\Lambda$CDM
model with parameters suggested by \textit{Planck}) and
GP-reconstructed $D_A(z)$ from cosmic chronometers (passively
evolving galaxies). We present more details on cosmic chronometers
method later in this section. In both cases the sample contained the
same objects. So the sample selected by the luminosity seems not to
be affected much by the cosmological model assumed. Parameter
fitting results are summarized in Table 1 and displayed in
Fig.~\ref{fig1}. The next issue is that our criterion seems uneven:
it cuts only one order of magnitude ($10^{27}$ W/Hz$<L<$ $10^{28}$
W/Hz) for intermediate-luminosity quasars, allowing the other
classes -- high and low-luminosity quasars -- to vary over many
orders of magnitude. In order to check for this, we also made
similar fits to high and low-luminosity samples limited to one order
of magnitude. Strictly speaking, we included in our analysis two
following samples: 1) low-luminosity and high-redshift quasars with
$10^{26}$ W/Hz$<L<$ $10^{27}$ W/Hz and $z>0.50$; 2) high-luminosity
and high-redshift quasars with $10^{28}$ W/Hz$<L<$ $10^{29}$ W/Hz
and $z>0.50$. The results shown in Table 1, tend to support the
conclusion that only intermediate-luminosity quasars exhibit no size
evolution and could be classified as standard rulers.
Fig.~\ref{fig2} shows the scatter plot of the three quasar
sub-samples in the ($\theta$, $z$, $S$) space of observable
quantities. One can see that while low-luminosity and
intermediate-luminosity samples are well separated, intermediate and
high-luminosity ones are partly mixed. This motivated us to perform
fits for combined intermediate plus high-luminosity samples with the
results displayed in respective rows of the Table 1. In conclusion:
one can see that quasars with intermediate-luminosities meet the
requirement for a standard rod: $|n|\simeq 10^{-3}$, $|\beta|\simeq
10^{-4}$. We remark that the linear size is dependent on the
rest-frame frequency, thus the zero value of $n$ should be checked
with multi-frequency VLBI data in the future. The best-fitted values
of $\beta$ and $n$ for this sub-population are significantly
different from the corresponding quantities for other sub-samples,
which supports the scheme of using a distinct strategy for treating
quasars with intermediate luminosities \footnote{We are
investigating the possibility that the distinction between
medium-luminosity quasars and high-luminosity ones has been
overstated. There is a degree of degeneracy between the parameters
$\beta$ and $n$, which arises because the luminosity $L$ and
redshift $z$ are correlated. Pending further investigations, we
prefer to leave the 31 high-luminosity quasars out of our
considerations. Low-luminosity quasars are mostly at low redshifts,
where the population is qualitatively different.}.

The redshift of intermediate-luminosity quasars ranges between
$z=0.462$ and $z=2.73$. In fact, the previous studies have given
plausible reasons for ignoring low-redshift quasars with $z<0.50$.
The high-redshift part ($z\geq0.50$) of the sample exhibits a
smaller dispersion in luminosity $L$, and the median angular size of
quasars in this range is nearly independent of redshift
\citep{Gurvits94,Jackson97}. In addition, there is no evidence of
any selection bias concerning radio luminosity \citep{Jackson04}
(weaker sources are distinctly smaller) when $z>0.50$.

\begin{figure*}
\begin{center}
\includegraphics[scale=0.6]{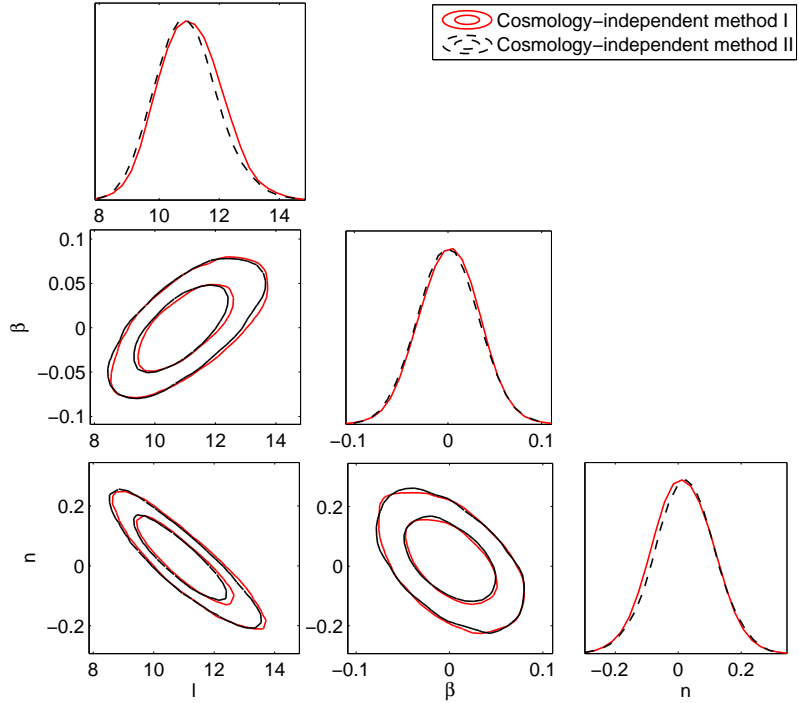}
\end{center}
\caption{ Constraints on the compact source parameters for
intermediate-luminosity quasars with the $D_A(z)$ function reconstructed form  $H(z)$ data. Red solid contours correspond
to the results with $H_0= 69.6 \pm 0.7 $ km sec$^{-1}$
Mpc$^{-1}$ (Cosmology-independent method I) and the black dashed
contours denote the results with $H_0= 73.24 \pm 1.74 $ km
sec$^{-1}$ Mpc$^{-1}$ (Cosmology-independent method II). Both of them
support the claim that the intermediate-luminosity quasars display
negligible dependence of their linear size parameter on the source
luminosity and redshift ($|n|\simeq 10^{-2}$ and $|\beta|\simeq
10^{-3}$). }
 \label{fig3}
\end{figure*}

Another issue which might be raised regarding our selection of
standard rods is the assumption of a particular cosmological model
--- $\Lambda$CDM with the most recent parameters
--- in the course our estimation of $l$, $\beta$, and $n$. Therefore, we
have undertaken a similar analysis with different cosmological
models and their best fitted parameters: WMAP9 observations in the
case of $\Lambda$CDM \citep{Hinshaw13}, as well as \textit{Planck}
observations in the case of XCDM and CPL parameterizations
\citep{Planck1}. The results we obtained were very similar to those
for the $\Lambda$CDM model best fitted to \textit{Planck} data.
However, this approach still depends on the cosmological model assumed.
Therefore, we also performed a model independent study. As
is well known, assuming the FRW metric of a flat universe,
the angular-diameter distance can be written as $D_A(z)= c/(1+z)
\int^{z}_{0}\frac{dz'}{H(z')}$, where $H(z)$ is the Hubble
parameter. Following the recent works of \citet{Wei16} inspired by
Gaussian processes (GPs) \citep{GP,Seikel2012}, we have reconstructed the $H(z)$ function from the recent Hubble parameter
measurements \citep{Zheng2016} and then derived $D_A$ covering the
redshift range of intermediate-luminosity quasars. Such approach is
independent of any specific cosmological model since the cosmic chronometers approach resulting in $H(z)$ measurements is independent of any assumptions about cosmology. Moreover, in order
to study the potential effect of the Hubble constant on the distance
reconstruction with this method, two
recent measurements of $H_{0}=69.6\pm0.7$ km $\rm s^{-1}$ $\rm
Mpc^{-1}$ \citep{Bennett14} and $H_{0}=73.24\pm1.74$ km $\rm s^{-1}$
$\rm Mpc^{-1}$ \citep{Riess16} were adopted in the GP method.
Distance reconstruction with the former prior on the Hubble constant is
denoted as ``Cosmology-independent method I", while
``Cosmology-independent method II" represents the distance
reconstruction with the latter prior on the Hubble constant. Constraint
results for the 120 intermediate-luminosity quasars with the two
model-independent methods are shown in Fig.~\ref{fig3} and also in Table 1. They
are consistent with those determined from the prior coming from the flat
$\Lambda$CDM model best fitted to \textit{Planck} data, hence our
results and discussions presented above are robust.

\begin{figure*}
\begin{center}
\includegraphics[scale=0.45]{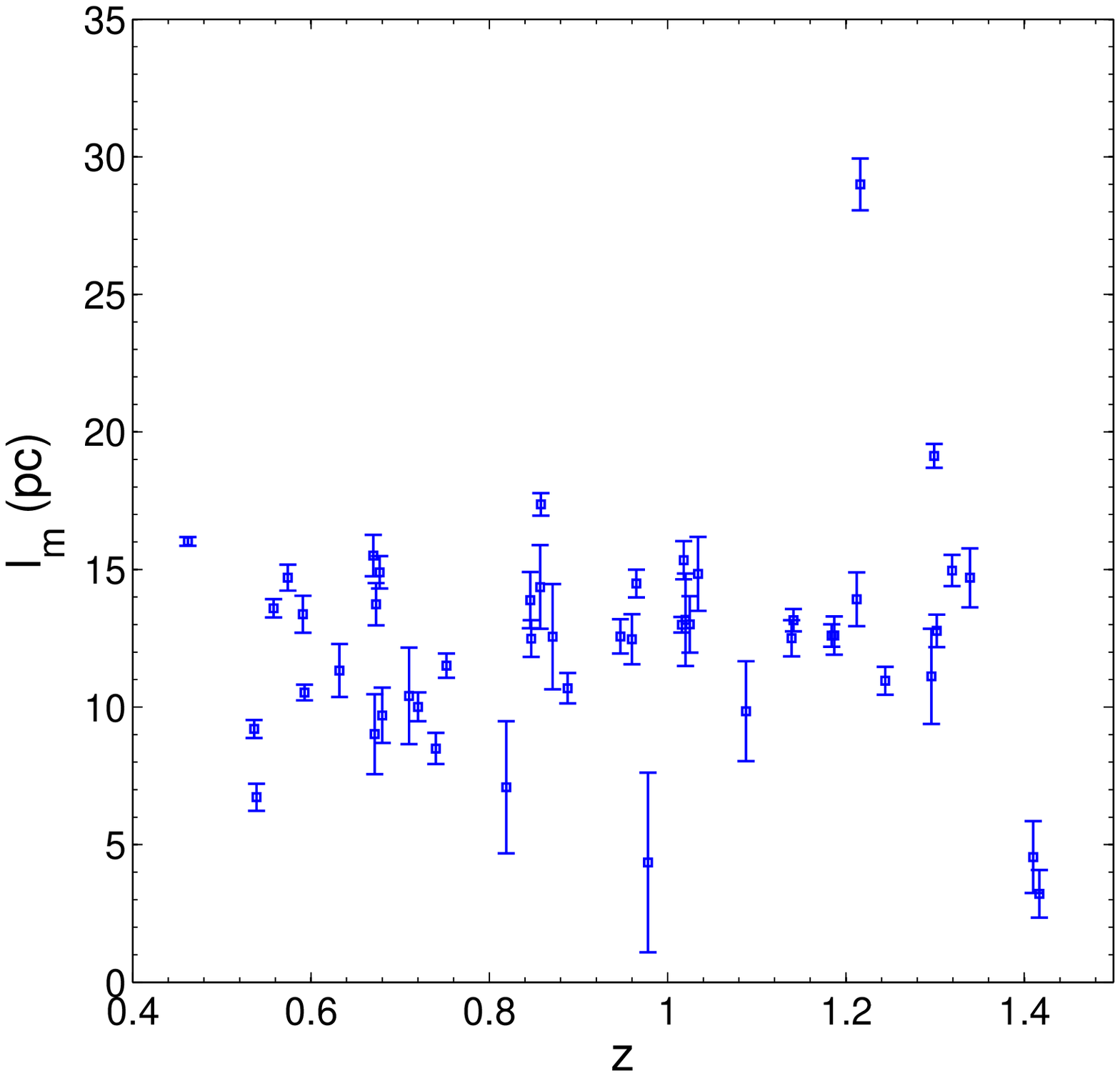}\includegraphics[scale=0.45]{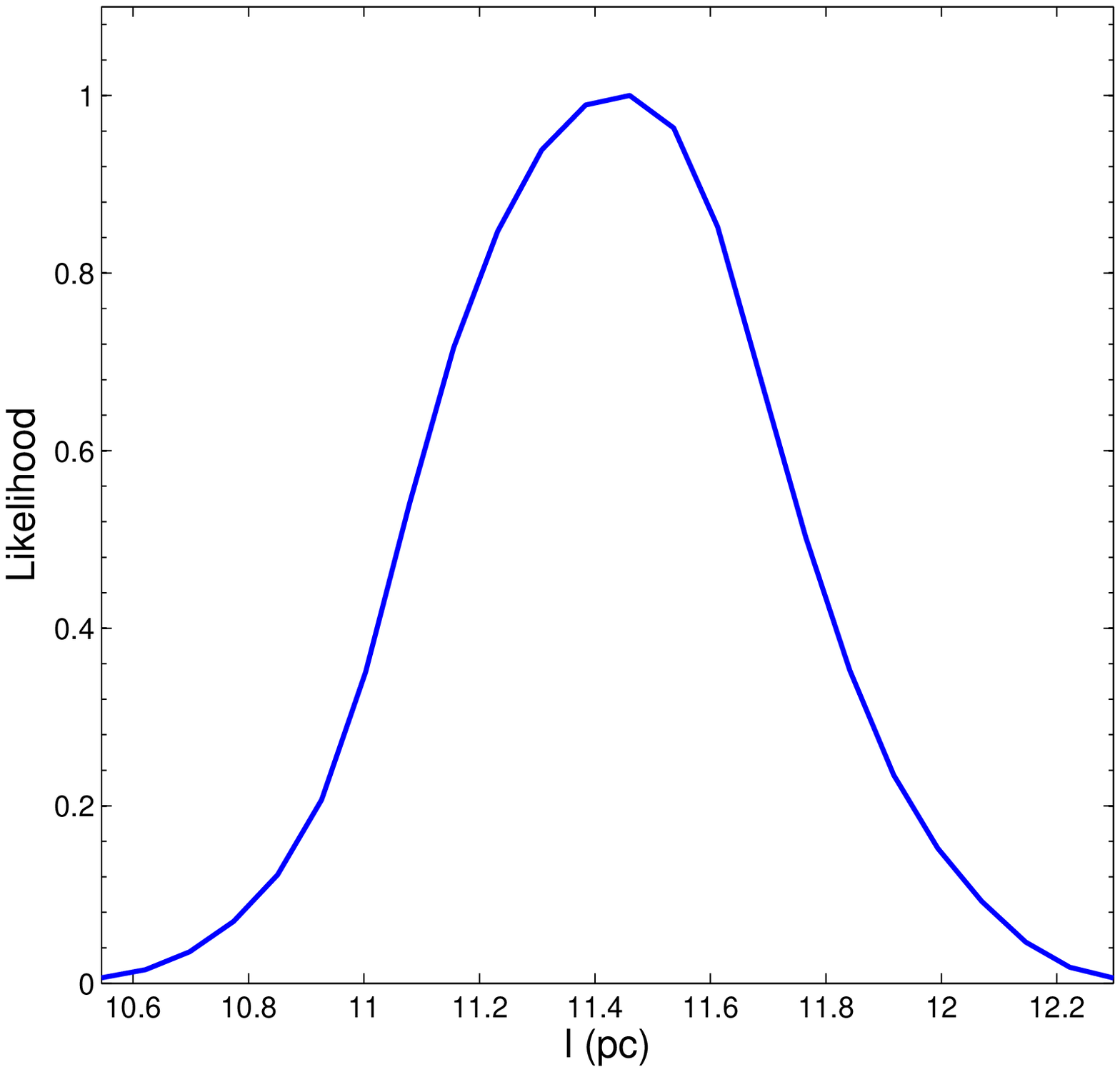}
\end{center}
\caption{ Left figure displays the intrinsic metric linear
size $l_m$ distribution of the 48 intermediate luminosity quasars in
the redshift space. Right figure shows the likelihood distribution
function of $l$ --  the corrected linear size of compact quasars.
Both figures are derived without explicity assuming the cosmological
model, but using SN Ia (Union2.1 sample). }\label{fig4}
\end{figure*}

We invoked a specific cosmological model in order to identify the
sub-sample which could then be used as standard rulers, and it
turned out that intermediate-luminosity quasars would serve this
purpose the best. Therefore in order to calibrate further (i.e. determine more accurately) the linear size of
the compact structure in the intermediate-luminosity radio quasars
we used the following method. Namely, the SN Ia
are commonly accepted standard candles in the Universe and from
their observed distance moduli we are able to recover luminosity
distances $D_L(z)$ covering the lower end of the quasar redshift
range: $z \lesssim 1.41$. By virtue of the reciprocity theorem, they
can be converted into angular diameter distances $D_A(z)$ and used
to calibrate the linear size of our milliarcsecond quasars. Applying
the redshift-selection criterion, $\Delta z=|z_{QSO}-z_{SN}|\leq
0.005$ (according to \citep{Cao11a}) to the Union2.1 compilation
\citep{Amanullah10}, we obtained 48 measurements of $D_A$ in the
supernova/quasar overlap region, corresponding to the quasar
redshifts. The intrinsic metric linear size $l_m$
distribution of the 48 intermediate luminosity quasars is also
shown in the left panel of Fig.~\ref{fig4}. Next we performed a similar
fitting procedure, but now allowing the quantity $l_m$ to have only
one free parameter, namely $l$, so that the values of $D_A$ inferred
from equation (\ref{lm}) match the supernova ones. As a result, we
obtained the following:
\begin{eqnarray}
&& l= 11.42\pm0.28 \ \mathrm{pc}. 
\end{eqnarray}
It should be noted that the 48 values of $D_A$ so derived
are not used in any other context; specifically they are not the
used directly in the preparation of the
angular-diameter-distance/redshift diagram shown on Fig.~\ref{fig6},
see discussion later. The likelihood function of $l$ is also shown
in Fig.~\ref{fig4}, from which one can see the perfect consistency
between the result derived from the cosmological-model-independent
analysis and that determined from theoretical cosmological
distances. In order to check the constraining power of quasars with
the corrected linear size, using the ``$\theta-z$" relation for the
full quasar sample, we were able to get stringent
constraints on both the matter density parameter
$\Omega_m=0.323^{+0.245}_{-0.145}$ and the Hubble constant
$h=0.663^{+0.070}_{-0.085}$ in the flat $\Lambda$CDM cosmology,
which are in reasonable agreement with those obtained from
\textit{Planck} observations. According to the unified
classification of Active Galactic Nuclei (AGN), 10 pc is the typical
radius at which AGN jets are apparently generated and there is
almost no stellar contribution \citep{Blandford78}. In the unified
scheme this radius defines a compact opaque parsec-scale core, which
is located between the broad-line region ($\sim$ 1 pc) and
narrow-line region ($\sim$ 100 pc) for QSOs. Recent AGN observations
\citep{Silverman09} indicated that, within 10 pc around the central
black hole, star formation rate is equal to the black hole's mass
accretion rate, which is consistent with the simulation results from
\citet{Hopkins10}.

It is instructive to compare the estimates of angular sizes used
here with those derived from more recent VLBI imaging observations
based on better uv-coverage. \citet{Pushkarev15} presented the
archival VLBI data of more than 3000 compact extragalactic radio
sources (hereafter called P15) observed at different frequencies,
$\nu=2\sim 43$ GHz. Using these angular-size measurements of radio
cores of active galactic nuclei (AGN) observed at 2 GHz, we can
estimate the linear sizes of $\sim 1600$ milliarcsecond sources
covering redshift range $z>0.46$, which provides a statistical value
for the characteristic linear size, $l=12.25$ pc ($H_0$=70 km
sec$^{-1}$ Mpc$^{-1}$). However, we should remark here that the P15
sample contains a wide class of extragalactic objects, including
quasars (belonging to different luminosity categories), radio
galaxies, and BL Lac objects, etc. Such measurement at 2 GHz gives a
weighted mean size, taken over all the radio sources, rather the
intermediate-luminosity quasars. In fact, 58 intermediate-luminosity
quasars included in the modified P85 sub-sample used in the study,
have also been observed by recent VLBI observations based on better
uv-coverage and included in the P15 sample. Using these angular-size
measurements of radio cores of intermediate-luminosity quasars
observed at 2 GHz, we estimate the statistical value for the
characteristic linear size, $l=11.51$pc, which is well consistent
with our 2.3 GHz result derived from the P85 sample.

\begin{figure*}
\begin{center}
\includegraphics[scale=0.6]{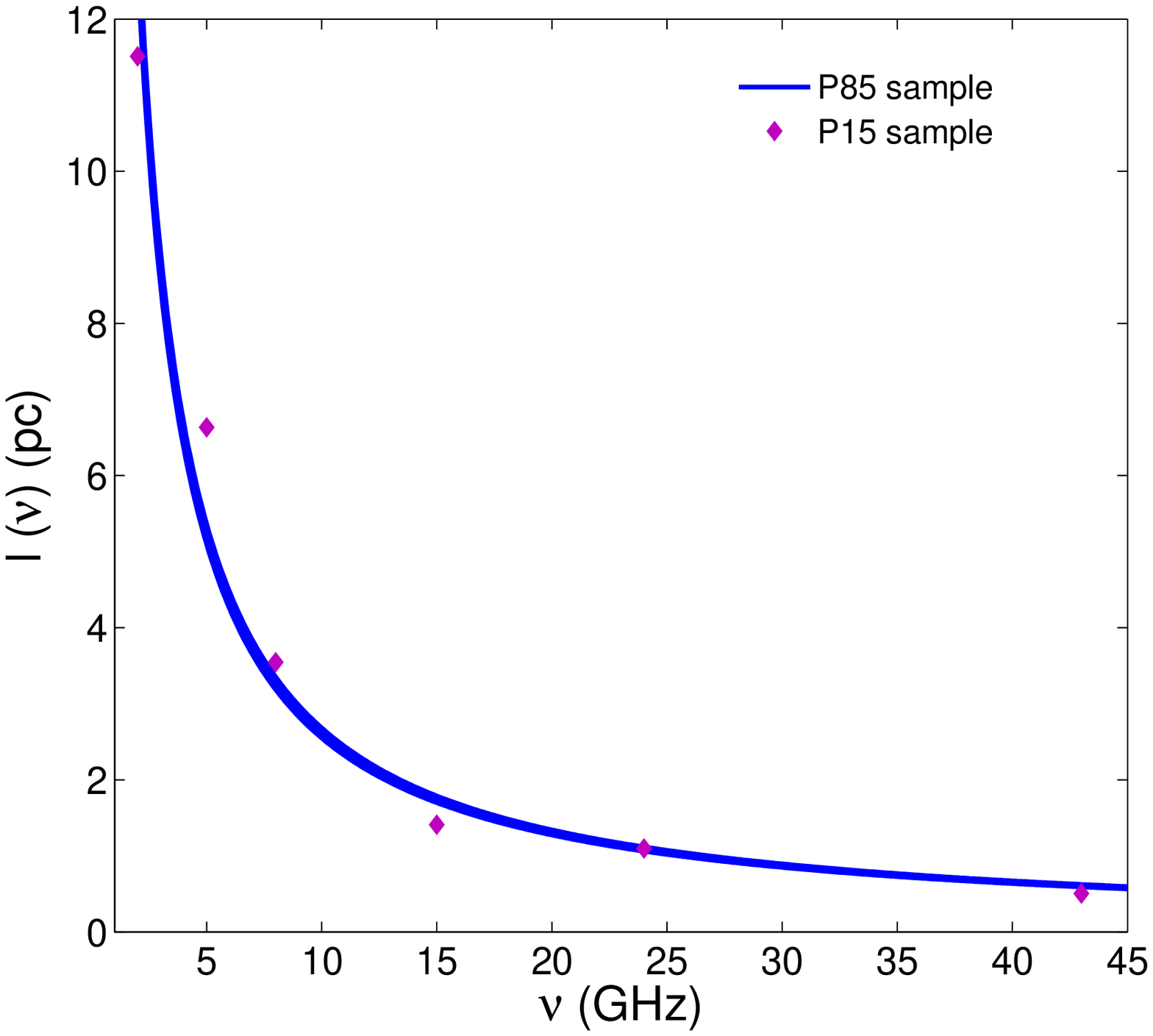}
\end{center}
\caption{ The plot of linear size of compact structure in
intermediate-luminosity quasars ($l$) versus frequency ($\nu$),
derived from both the P85 sample and P15 sample. BK 79 jet model
based on the linear-size measurement from the P85 sample ($l=
11.42\pm0.28$ pc at 2.29 GHz) is also plotted. }\label{fig5}
\end{figure*}

\begin{figure*}
\begin{center}
\includegraphics[scale=0.6]{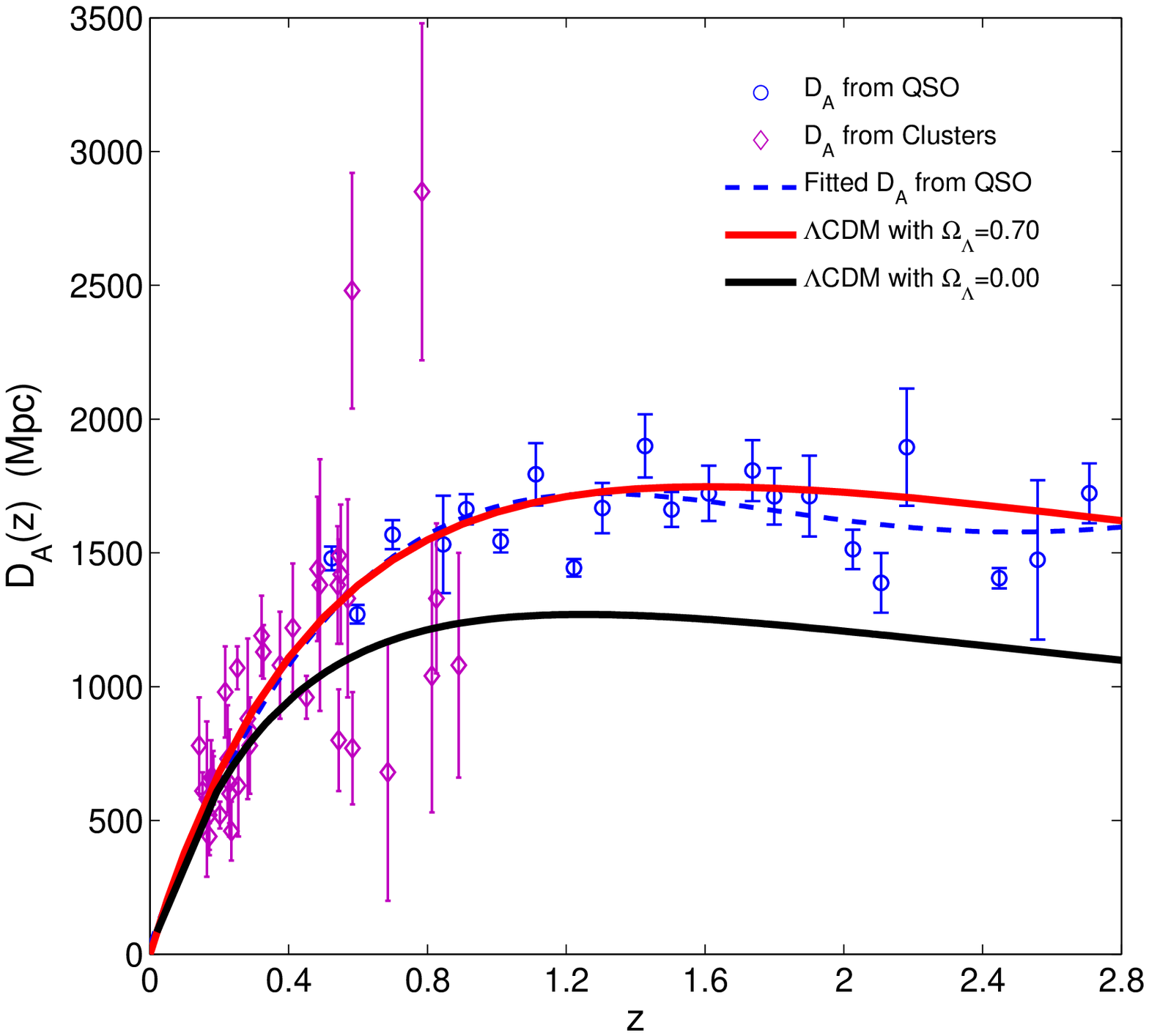}
\end{center}
\caption{ Angular diameter distances $D_A(z)$ estimated from quasars
as standard rulers (blue circles). Angular diameter distances from
galaxy clusters (purple diamonds) are also added for comparison.
Theoretical predictions of $\Lambda$CDM models with
$\Omega_\Lambda=0.00$ and $\Omega_\Lambda=0.70$ are denoted by black
and red solid lines, respectively. }\label{fig6}
\end{figure*}

In the conical jet model proposed by \citet{Blandford79} [hereafter
BK79], if we observe a given millarcsecond source at different
frequencies $\nu$, its observed size falls as the frequency
increases, being proportional to $\nu^{-1}$
\citep{Marscher80,O'Sullivan09}. Therefore, VLBI results obtained at
other frequencies on the same sources as used in the current study
should be used for verification of the results. Based on the
multi-frequency angular size measurements from the P15 sample, we
estimate characteristic linear sizes as $l=6.63$ pc at 5 GHz,
$l=3.55$ pc at 8 GHz, $l=1.41$ pc at 15 GHz, $l=1.10$ pc at 24 GHz,
and $l=0.50$ pc at 43 GHz. These figures are compatible with the
expected $\nu^{-1}$ behaviour based on the measurement from the P85
sample (See Fig.~\ref{fig5}).

Next we attempted to construct an empirical relation $D_A(z)$
extending to higher redshifts on the basis of individual quasar
angular sizes, using Eq.~(\ref{theta}); we obtained the $D_A$
measurement and  corresponding uncertainty for each quasar. However,
this procedure  results in large uncertainties in $D_A$ -- a problem which
has been encountered previously \citep{Gurvits94,Gurvits99},
as can be seen from plots of the measured angular size  against
redshift therein. This problem remains even after 13 systems with
very large ($\sim 50\%$) uncertainties are removed. In order to
minimize its influence on our analysis, we have chosen to bin the
remaining 107 data points and to examine the change in $D_A$ with
redshift. The final sample was grouped into 20 redshift bins of
width $\Delta z=0.10$. Fig.~\ref{fig6} shows the median values of
$D_A$ for each bin plotted against the central redshift of the bin.
For comparison, the two curves plotted as solid lines represent
theoretical expectations from the concordance flat $\Lambda$CDM model and
the Einstein - de Sitter model. One can see that the latter is disfavored at high
confidence,  more precisely with $\chi^2 /d.of. = 6.28$ which corresponds to $3\sigma$ confidence level. More
importantly, the angular diameter distance information obtained from
quasars has helped us to bridge the ``redshift desert'' and extend
our investigation of dark energy to much higher redshifts.

\begin{figure*}
\begin{center}
 \includegraphics[scale=0.42]{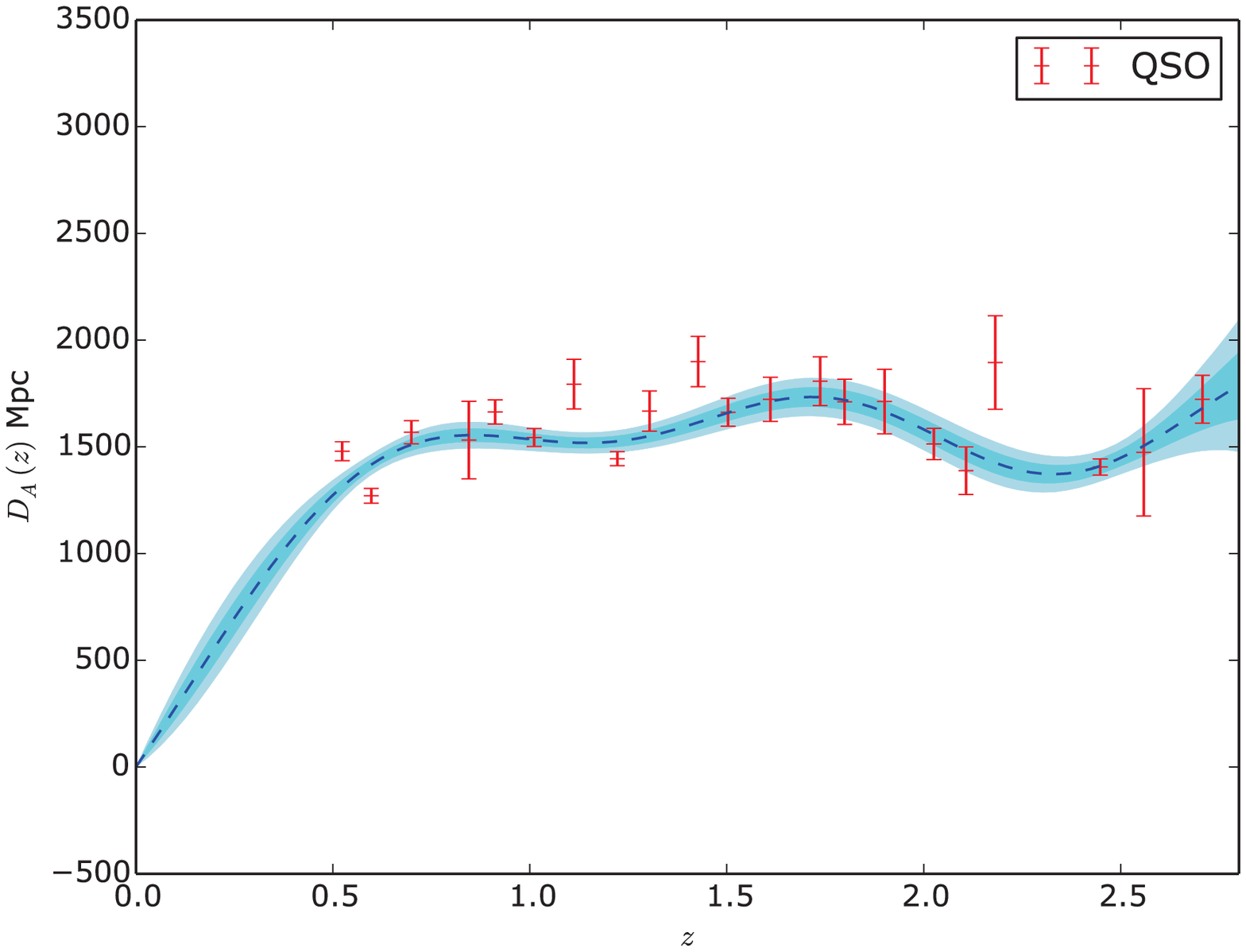}\includegraphics[scale=0.42]{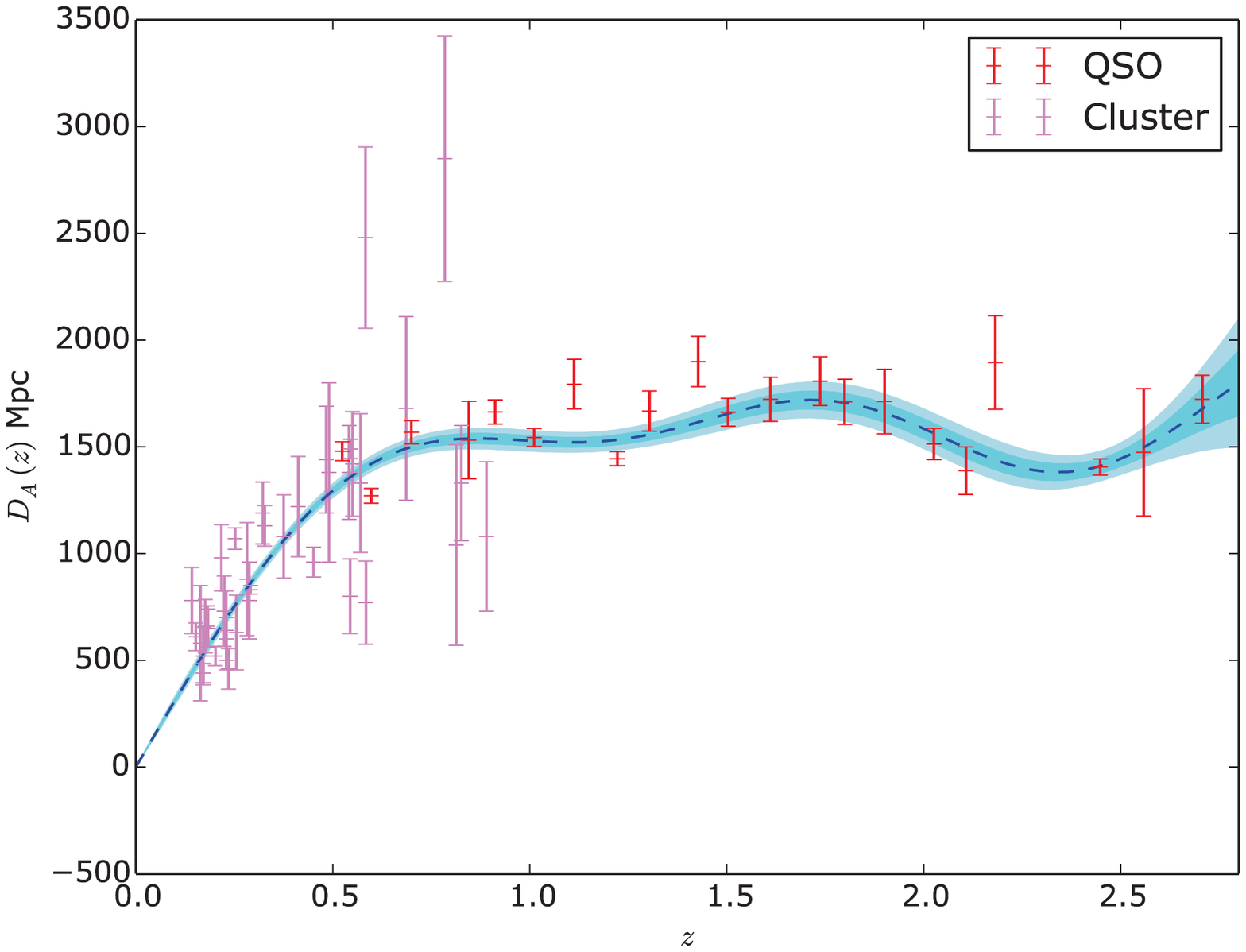} \\
 \includegraphics[scale=0.42]{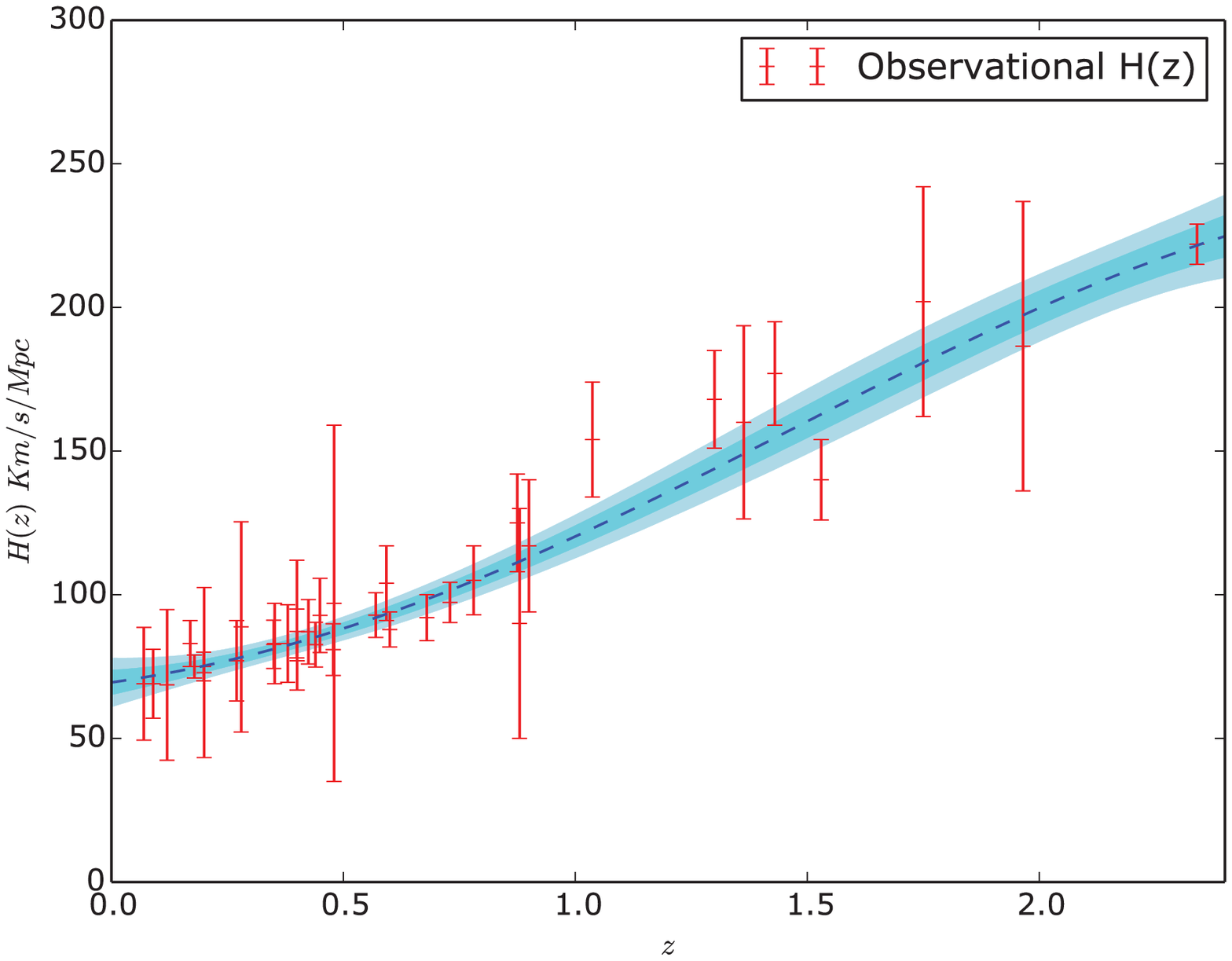}\includegraphics[scale=0.42]{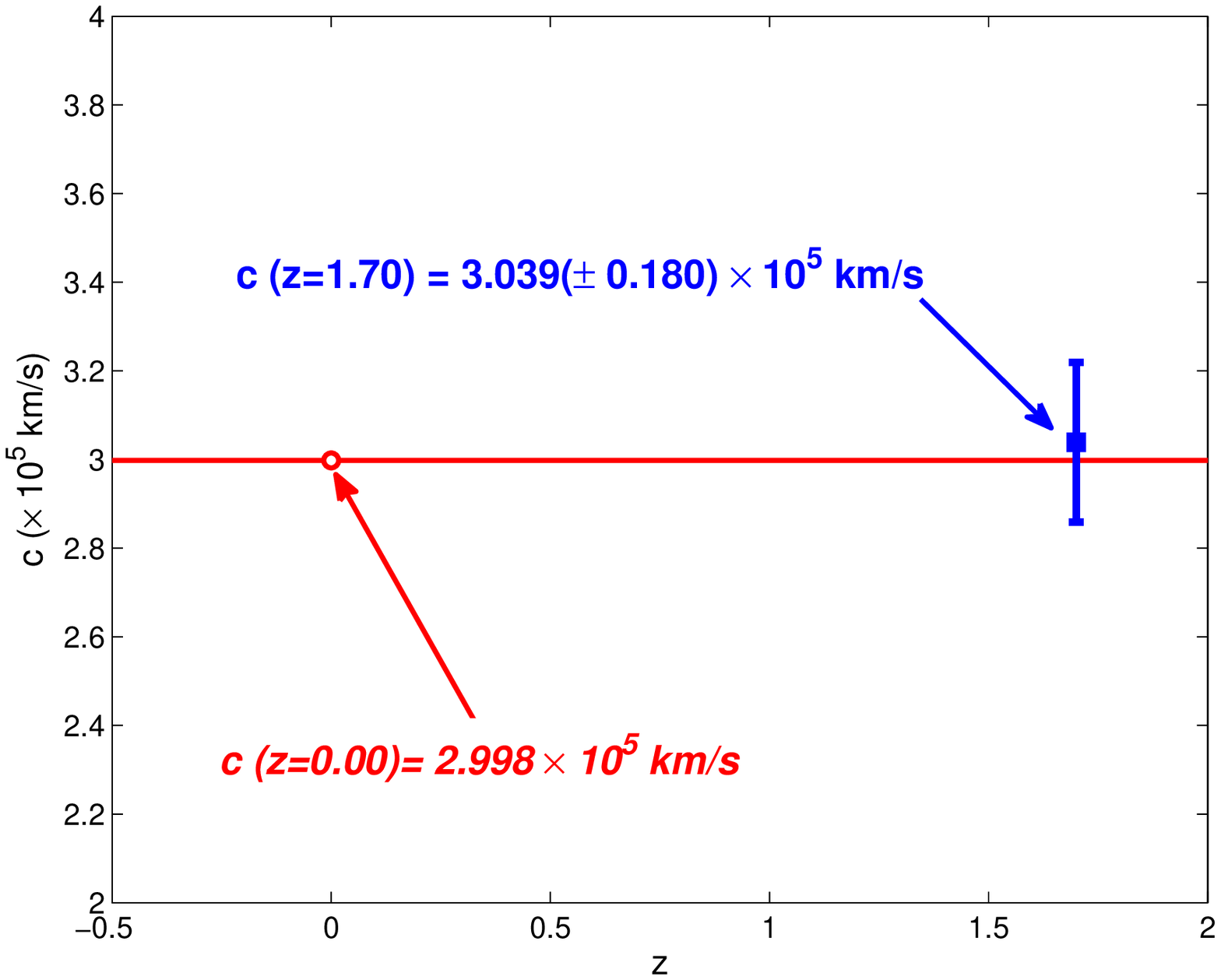}
\end{center}
\caption{Upper left: GP reconstructed $D_{A}(z)$ using the data
concerning quasars;  Upper right: GP reconstructed $D_{A}(z)$ using
the data with both quasar and cluster observations; Lower left: GP
reconstructed $H(z)$ data with Hubble parameter observations; Lower
right: The measured speed of light $c$ at redshift $z=1.70$ (blue
square with error bars) and $c_0$ at redshift $z=0$ (red circle).
Standard assumes that $c$ is a universal constant $c_{0}$ so the red
horizontal line has also been added for comparison. } \label{fig7}
\end{figure*}

Finally, by combining the measurements of $D_{A}$ and the Hubble
parameter $H(z)$ at different redshifts, it should be possible to
measure the speed of light $c$. For this purpose one must measure
the redshift $z_m$ at which $D_A(z)$ reaches its maximum, then
measure the angular diameter distance and the expansion rate at this
particular redshift, and finally evaluate the speed of light through $c=D_{A}(z_m)H(z_m)$.
Even though, from a purely theoretical point of view this idea is
appealing (provided one is indeed able to measure $D_A(z)$ and
$H(z)$ accurately and in the absolute sense) it turns out to be very
difficult from the observational point of view. Therefore the papers
\citep{Salzano15,Salzano16} by the proponents of this method refer
only to simulated (fictitious) data. Our attempt is the first one
performed on real data for this purpose.

Our quasar sample is sufficient to reconstruct the profile of
$D_A(z)$ up to the redshifts $z\sim 3$. However, large uncertainties
of the $D_{A}$ measurements combined with the intrinsically large
plateau about $z_{m}$ make it impossible to determine $z_m$
precisely. On the other hand, one can use a powerful reconstruction
method \citep{GP,Seikel2012} based on Gaussian Processes (GPs). We
have employed these to reconstruct the function $D_{A}(z)$ in order
to find $z_{m}$, and also to reconstruct $H(z)$ function to find
$H(z_{m})$. The data on the $H(z)$ function we used for the
reconstruction, were acquired by means of two different techniques.
The first was based on the so called cosmic chronometers
\citep{JimenezLoeb}, i.e. the differential ages of massive,
early-type galaxies evolving passively on a timescale longer than
their age difference. Recent analysis of the Baryon Oscillation
Spectroscopic Survey (BOSS) Data Release by \citet{Moresco16}
enlarged the previous data set \citep{Ding2015,Moresco15} to a total
number of 30 measurements of $H(z)$. The second technique relied on
the analysis of BAO and provided us 6 further, precise measurements
of $H(z)$ at 6 different redshifts (see Table 1 of \citet{Zheng2016}
for details and appropriate references to the original sources).

The $D_{A}(z)$ function reconstructed from the binned quasar data is
shown in Fig.~\ref{fig7} (upper left panel).
From this reconstructed function we obtained the redshift at which
$D_A(z)$ reaches its maximum. The value is $z_m=1.70$ and the
corresponding angular diameter distance $D_A(z_m)=1719.01\pm43.46$
Mpc. We have also enhanced the $D_A$ measurements with 38 galaxy
clusters \citep{Boname06} and repeated the reconstruction of
$D_A(z)$ (see upper right panel of the Fig.~\ref{fig7}). However,
because of the redshift range of galaxy clusters ($z=0.16\sim
0.89$), their effect on the determination of $z_m$ turned out to be
negligible. The lower left panel of the Fig.~\ref{fig7} shows the
reconstructed $H(z)$ function based on 36 Hubble parameter
measurements. From this reconstructed relation we obtained
$H(z_m)=176.77\pm6.11$ Mpc/km/s at the maximum redshift. Therefore,
our final assessment of the speed of light at $z_m$ is the following
$c=3.039(\pm0.180)\times 10^5$ km/s. This is the first measurement
of the speed of light in a cosmological setting referring to a
distant past (at redshift $z_m$ the Universe was only $3.80$ Gyr
old). However, the result is in perfect agreement with the value
$c_{0} \equiv 2.998\times 10^5$ km/s measured ``here and now'',
which according to standard physics should be a universal constant
of Nature. However, the agreement is only with our best fit central
value, and the (inevitable) uncertainty in our result means that
some window of opportunity is still open for exotic
varying-speed-of-light scenarios. There are many ways in which such
a constraint might be improved using our technique, which has paved
one particular way in which one can derive the speed of light in the
past. It should be stressed that our analysis and the assessment of
$c$ were based on the radio source data from very old VLBI surveys.
The prospect of future multi-frequency VLBI surveys, comprising much
more sources with higher sensitivity and angular resolution,
especially at redshifts around $\approx 1.5-1.6$ where the redshift
of $D_A(z)$ maximum is very likely to be located, means that our
constraints could be further improved by an order of magnitude.
Moreover, one can imagine that a better reconstruction method might
be devised, to process the raw radio quasar data at a preliminary
stage. Hopefully the measurement of $c$ from quasars might be
considerably improved by using future powerful probes such as BAO
data from a survey like \textit{WFIRST} \citep{WFIRST,Antonucci93}.
Finally, we make a remark on the determination of the characteristic
fixed length of our quasars.  The distance moduli, used to obtain
the 48 measurements of $D_A(z)$ in the supernova/quasar overlap
region, depend upon absolute magnitudes given in \citet{Amanullah10}
(Table 10), which in turn depend upon an assumed value of Hubble's
constant; the latter is given as $H_0=70$ km sec$^{-1}$ Mpc$^{-1}$.
Our values of $D_A(z)$ scale as $h^{-1}$, as do the corresponding
values of linear size.  Hence our particular value should written as
$l=11.42\times(0.70/h)=7.99 h^{-1} $ pc.

\section*{Acknowledgements}

The authors would like to thank the referee for constructive comments, which allowed to improve the manuscript substantially.
We are grateful to Zhengxiang Li, Yang Chen, and Meng Yao for
helpful discussions. This work was supported by the Ministry of
Science and Technology National Basic Science Program (Project 973)
under Grants Nos. 2012CB821804 and 2014CB845806, the Strategic
Priority Research Program ``The Emergence of Cosmological Structure"
of the Chinese Academy of Sciences (No. XDB09000000), the National
Natural Science Foundation of China under Grants Nos. 11503001,
11373014 and 11073005, the Fundamental Research Funds for the
Central Universities and Scientific Research Foundation of Beijing
Normal University, China Postdoctoral Science Foundation under grant
No. 2015T80052, and the Opening Project of Key Laboratory of
Computational Astrophysics, National Astronomical Observatories,
Chinese Academy of Sciences. Part of the research was conducted
within the scope of the HECOLS International Associated Laboratory,
supported in part by the Polish NCN grant DEC-2013/08/M/ST9/00664 -
4. M.B. gratefully acknowledges this support. This research was also
partly supported by the Poland-China Scientific \& Technological
Cooperation Committee Project No. 35-4. M.B. obtained approval of
the foreign talent introducing project in China and gained special
funding support from the foreign knowledge introducing project.

\end{document}